\documentclass[twocolumn,aps,preprintnumbers,prl,showpacs,amsfonts,amsmath,amssymb,superscriptaddress,floatfix]{revtex4}

\usepackage{graphicx}% Include figure files
\usepackage{dcolumn}% Align table columns on decimal point
\usepackage{bm}% bold math

\newcommand \bs{\begin{subequations}}
\newcommand \es{\end{subequations}}
\newcommand \bea{\begin{eqnarray}}
\newcommand \eea{\end{eqnarray}}
\newcommand \be{\begin{equation}}
\newcommand \ee{\end{equation}}

\begin{document}

\title{Observation of spatial quantum correlations induced by multiple scattering of non-classical light}
\author{S. Smolka}\email{stsm@fotonik.dtu.dk}
\affiliation{DTU Fotonik, Department of Photonics Engineering, Technical University of Denmark, Building 345V, 2800
Kgs. Lyngby, Denmark}
\author{A. Huck}
\affiliation{DTU Physics, Department of Physics, Technical University of Denmark, Building 309, 2800 Kgs. Lyngby, Denmark}
\author{U.L.\ Andersen}
\affiliation{DTU Physics, Department of Physics, Technical University of Denmark, Building 309, 2800 Kgs. Lyngby, Denmark}
\author{A. Lagendijk}
\affiliation{FOM Institute for Atomic and Molecular Physics, Kruislaan 407, 1098 SJ Amsterdam, The Netherlands}
\author{P. Lodahl}\email{pelo@fotonik.dtu.dk}\homepage{http://www.fotonik.dtu.dk/quantumphotonics}
\affiliation{DTU Fotonik, Department of Photonics Engineering, Technical University of Denmark, Building 345V, 2800
Kgs. Lyngby, Denmark}

%\date{\today}

\begin{abstract}
We present the experimental realization of spatial quantum
correlations of photons that are induced by multiple scattering of
squeezed light. The quantum correlation relates photons
propagating along two different light trajectories through the
random medium and is infinite in range. Both positive and negative
spatial quantum correlations are observed when varying the quantum
state incident to the multiple scattering medium, and the
magnitude of the correlations is controlled by the number of
photons. The experimental results are in excellent agreement with
recent theoretical proposals by implementing the full quantum
model of multiple scattering.
\end{abstract}

 \maketitle

\begin{figure}[ht]
  \centering
  \includegraphics[width=0.5\textwidth]{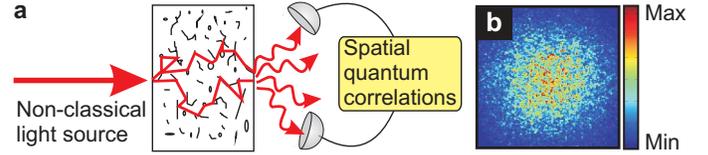}
\caption{(color online). (a) Illustration of different light
trajectories in a multiple scattering medium that are correlated
 depending on the quantum state of the light incident to the medium. (b) Measured spatial intensity distribution of light transmitted through a
multiple scattering medium displaying a volume speckle pattern.}
  \label{fig01}
\end{figure}

Multiple scattering of light waves propagating through a
disordered medium can be described as a random walk of the
direction of propagation \cite{nature453p495}, whereby the medium
becomes opaque \cite{prl101p120601}. Such random scattering
determines the performance of nanophotonic devices based on, e.g.,
photonic crystals \cite{prl94p033903} or surface plasmon
polaritons \cite{prl89p186801}, and can be applied for enhancing
the information capacity for optical communication
\cite{science287p287}. In multiple scattering, the different
possible light trajectories through the random medium will
interfere and a complex spatial intensity distribution of light is
generated, cf. Fig.~\ref{fig01}(b). Usually interference effects
tend to vanish after averaging over all possible realizations of
the disorder and the light transport is described by diffusion
theory. Under conditions where multiple scattering is strong,
light diffusion is modified leading to spatial correlations such
as mesoscopic fluctuations \cite{revmodphys71p313,prl81p5800,prl64p2787}. These
correlations imply that the light intensity observed at one
position after the medium depends on the intensity at a different
position even after averaging over all ensembles of disorder.

The significance of the quantum nature of light in a multiple
scattering context has only been addressed recently
\cite{prl81p1829,prl89p043902,prl95p173901,prl94p153905,prl97p103901,optexpress14p6919}.
Pioneering work has provided the theoretical framework for a
quantum optical description of multiple scattering
\cite{prl81p1829} that was used to predict the existence of
spatial quantum correlations induced by multiple scattering of
non-classical light \cite{prl95p173901}. These quantum
correlations exist even in the diffusive regime of multiple
scattering and add to the traditional mesoscopic correlations that
are of classical origin \cite{revmodphys71p313}. Here we present
the experimental demonstration of spatial quantum correlations
induced by multiple scattering of squeezed light.

Due to the Heisenberg uncertainty principle intrinsic quantum
fluctuations are always present. Although always of inherent
quantum origin, photon fluctuations are classified as either
classical or non-classical if similar fluctuations can be induced
by classical light sources or not. The classical boundary
corresponds to Poissonian photon statistics where the variance of
the photon number fluctuations equals the mean number of photons.
A purely quantum regime exists where photon fluctuations are
reduced beyond the classical limit leading to sub-Poissonian
photon statistics.

The number of photons exiting a multiple scattering medium in a
specific direction can be correlated or anti-correlated with the
number of photons in another direction, and this spatial
correlation depends on the quantum state of light illuminating the
medium, c.f. Fig~\ref{fig01}(a). The degree of correlation is
quantified by the spatial quantum correlation function \cite{note}
\begin{equation} \overline{C_{abab'}^{Q}} = \frac{\overline{\left<\hat{n}_{ab}
\hat{n}_{ab'} \right>}}{\overline{\left<\hat{n}_{ab}\right>}
\times \overline{ \left<\hat{n}_{ab'} \right>}}-1, \label{eq01}
\end{equation}
where $a$ labels the direction of the incident light and $b$ and
$b'$ the two different output directions. The operator
$\hat{n}_{ab}$ represents the number of photons in output mode
$b$. $\left<\ldots \right>$ denotes the quantum mechanical
expectation value while the bars refer to an average over all
realizations of disorder. The need for two averages is due to the
fact that quantum fluctuations and random multiple scattering are
both stochastic processes. $\overline{C_{abab'}^{Q}}$ is the
quantum optical generalization of the classical intensity
correlation function that has been intensely investigated in
multiple scattering experiments \cite{revmodphys71p313}. Note that
the quantum correlations are induced despite the fact that the
response of the multiple scattering medium is fully linear. Thus,
a non-classical resource combined with linear optics is quite
generally sufficient for many quantum optics experiments, and even
enable quantum computing \cite{nature409p46}.

The spatial quantum correlation function defined in
Eq.~\eqref{eq01} depends sensitively on the number and
fluctuations of photons incident on the sample through direction
$a$. Restricting to a theory for discrete spatial and temporal
degrees of freedom, it is given by
\begin{equation} \overline{C_{abab'}^{Q}}= \left[1 + \frac{F_a-1}{\left< \hat{n}_a
\right>} \right] \left[1 + \overline{C_{abab'}^{(2)}} +
\overline{C_{abab'}^{(3)}} \right] -1. \label{eq02}
\end{equation}
$\left< \hat{n}_a \right>$ is the average number of incident
photons and the Fano factor $F_a = \Delta n_a^2/\left< \hat{n}_a
\right>$ gauges the variance in the photon number fluctuations
specified within the spectral bandwidth relevant for the
measurement. $\overline{C^{(2)}}$ and $\overline{C^{(3)}}$ are
classical spatial correlation functions \cite{revmodphys71p313}
that can be neglected in the diffusive regime of multiple
scattering that is of concern here, but it would certainly be
exciting to study their quantum optical equivalents as well. The
size of the Fano factor determines the transition from the
classical $(F_a\ge1)$ to the non-classical regime $(F_a < 1).$
Using non-classical light opens the door to a genuine quantum
regime where $-1 \le \overline{C_{abab'}^{Q}} < 0,$ corresponding
to spatially anti-correlated photons. Such spatial quantum
correlations are of \emph{infinite range} in the sense that the
magnitude is independent of the angular difference of the two
output directions, which translates into a spatial separation in
the far field. This is an example of the fundamentally new
phenomena that arise in quantum optical descriptions of multiple
light scattering. So far, quantum optical variables have been
measured in very few multiple scattering experiments that only
have focused on the regime of classical photon fluctuations
\cite{prl94p153905,prl97p103901}, while spatial quantum
correlations were not demonstrated.

The spatial quantum correlation of Eq.~\eqref{eq01} can be
obtained by recording the photon fluctuations of the total
transmission $\overline{\Delta {n}_T^2}$. They are related through
\cite{prl95p173901}
\begin{equation} \overline{\Delta {n}_T^2} = \frac{\ell}{L} \left<\hat{n}_a\right> + \frac{\ell^2}{L^2} \left<\hat{n}_a \right>^2
\overline{C_{abab'}^{Q}}, \label{eq03}\end{equation}
which holds in the diffusive regime of multiple scattering where
the spatial correlation function is independent of the output
direction $b$. A similar equation applies to the case of total
reflection measurements. Thus, the spatial quantum correlation
function $\overline{C_{abab'}^{Q}}$ is directly accessible by
measuring in addition to the photon fluctuations also the average
number of photons entering the sample $\left<\hat{n}_a\right>$ and
the ratio of the transport mean free path to the sample thickness
$\ell/L.$ Note that the latter ratio is modified by so-called
extrapolation lengths in order to account for the effects of the
interface of the multiple scattering medium \cite{Rivas99}. In the
experiment we extract the spatial correlation function in the
frequency domain, which has dimensions of an inverse bandwidth.
The complete continuous mode theory will be described elsewhere
\cite{Smolka-prep}.
\begin{figure}[ht]
  \centering
  \includegraphics[width=0.5\textwidth]{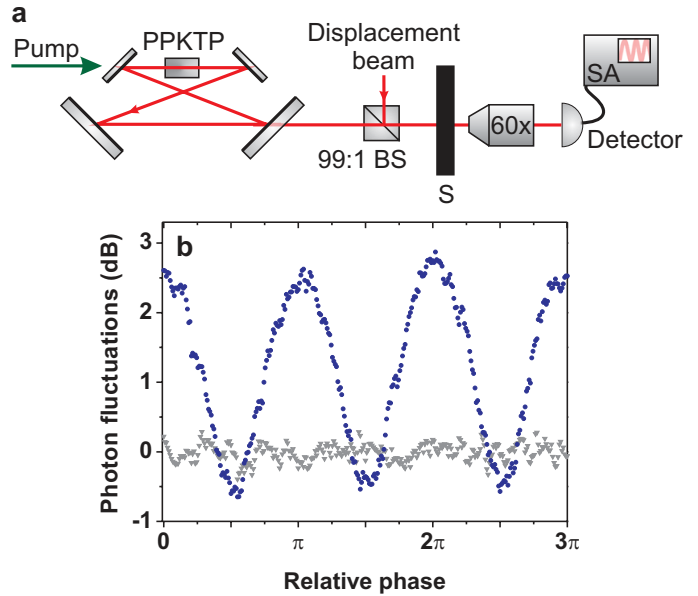}
\caption{(color online). (a) Sketch of the experimental setup.
Vacuum squeezing is generated in an optical parametric oscillator
with a PPKTP nonlinear crystal, overlapped with a displacement
beam on a beam splitter (BS), and directed onto the sample (S).
The multiple scattered light is imaged onto a photo-detector and
its photon fluctuations are recorded using an electronic spectrum
analyzer (SA). A similar setup is used for reflection measurements
illuminating the back side of the sample.
(b) Measured photon fluctuations after multiple scattering
recorded in reflection geometry ($L=20\:\mu\text{m}$). The photon
fluctuations of the non-classical light source (blue circles) are
recorded relative to the photon fluctuations associated with the
classical limit (gray triangles). Depending on the phase of a
displacement beam, photon fluctuations below or above the
classical limit are detected. The detection frequency was
$3.93\:\text{MHz}$ and the resolution bandwidth $300\:\text{kHz}$.
}
  \label{fig02}
\end{figure}

As a non-classical resource we use squeezed light generated in a
second-order nonlinear process at a wavelength of
$\lambda=1064\:\text{nm}$ \cite{prl57p2520}. This versatile source
features continuous tuning between classical (super-Poissonian)
and non-classical (sub-Poissonian) photon fluctuations by varying
the relative phase between the squeezed vacuum state generated in
the nonlinear process and a bright displacement beam. We are able
to reduce photon fluctuations ($F_a=0.52 \pm 0.02$) below the
classical limit $(F_a = 1)$ and induce excess fluctuations above
the classical limit $(F_a = 4.6\pm  0.2)$. The multiple scattering
experiment is conducted by focusing the non-classical light source
onto either the front surface or the back surface of the sample,
cf. Fig.~\ref{fig02}(a), to perform total transmission and total
reflection measurements, respectively. We record the total number
of photons and the photon fluctuations transmitted through or
reflected from the sample. The multiple scattering samples consist
of TiO$_2$ that has been grounded into strongly scattering
particles with a typical size of $200\:\text{nm}$. A range of
samples with different thicknesses has been fabricated, and the
total transmission of each sample was recorded with an integrating
sphere. From these measurements the transport mean free path is
found to be $\ell = (0.9 \pm 0.3)\:\mu\text{m}$ and all samples
are in the diffusive regime where $\lambda/2 \pi < \ell \ll L.$

Figure~\ref{fig02}(b) shows the measured photon fluctuations,
recorded in reflection, for squeezed light illumination of the
multiple scattering medium. The classical limit was recorded by
blocking the squeezed beam. The phase of the displacement beam is
scanned in order to continuously tune the photon fluctuations of
the incident light below and above the classical limit. We record
photon fluctuations that are reduced $0.5 \pm 0.1\:\text{dB}$
below the classical limit after the photons have been multiple
scattered, which corresponds to a Fano factor for the total
reflection of $F_R = 0.90 \pm 0.02$. Correcting for the overall
detection efficiency of $(37 \pm  1) \%$, we infer $F_R = 0.72 \pm
0.04$ for light exiting the multiple scattering sample. This
reduction is the direct experimental proof that non-classical
properties of light survive the complex stochastic process of
multiple scattering even after ensemble averaging over
realizations of disorder despite the common belief that quantum
properties of light are fragile. We calculate the average number
of scattering events in the multiple scattering process
\cite{prl60p1134}, which is given by $N \sim (L/\ell)^2$. For the
particular sample used for the measurements of Fig.~\ref{fig02}(b)
we get $N \approx 1000$, including effects of the interface, which
quantifies the complexity of the multiple scattering process. The
observed reduction in the photon fluctuations shows that the
information capacity associated with the multiple scattering
channels can be enhanced beyond the classical limit, as predicted
theoretically \cite{prl89p043902}.
\begin{figure}[ht]
  \centering
  \includegraphics[width=0.5\textwidth]{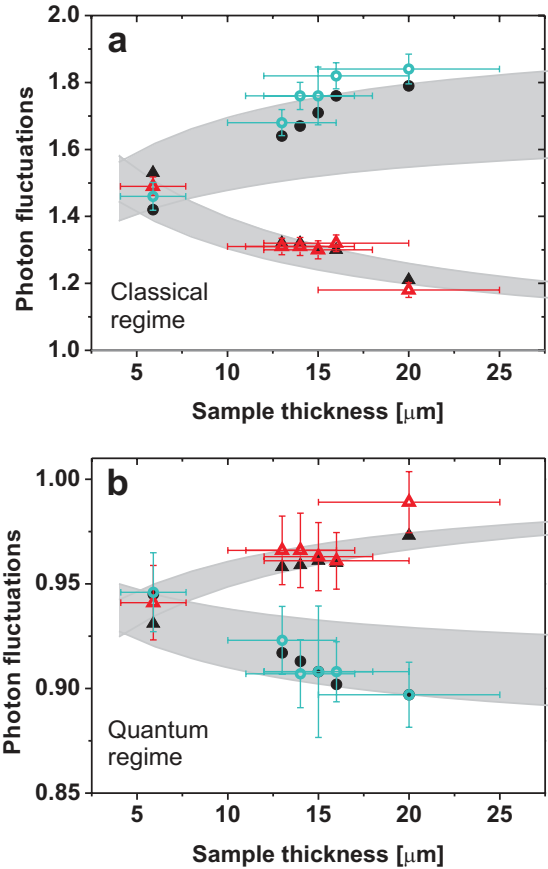}
\caption{(color online). Measured transmitted (red triangles) and
reflected (blue circles) photon fluctuations after multiple
scattering of light with classical (a) and non-classical (b)
photon fluctuations versus sample thickness. The data are
presented relative to the classical limit, which marks the
boundary between the classical regime and the quantum regime.
Every data point is obtained after ensemble averaging over six
different sample positions. The black data points are comparison
to theory taking into account the collection efficiency of the
detection setup, which was measured separately. The collection
efficiency varies for each measurement point since it depends on
sample thickness. The boundaries of the shaded areas correspond to
the theoretical predictions for minimum and maximum efficiencies.
}
  \label{fig03}
\end{figure}

We have carried out a detailed investigation of the transport of
non-classical and classical photon fluctuations through the
multiple scattering medium for a range of different sample
thicknesses. Figure~\ref{fig03} displays the detected photon
fluctuations after multiple scattering relative to the
fluctuations associated with the classical limit, which are
plotted versus the sample thickness. Using light sources with
classical fluctuations, the multiple scattered light always
displays excess photon fluctuations corresponding to the classical
regime (Fig.~\ref{fig03}(a)). Non-classical light allows entering
the quantum regime where the photon fluctuations are reduced below
the classical limit, see Fig.~\ref{fig03}(b). Our experimental
results can be compared to the predictions from the full quantum
theory for multiple scattering of photons
\cite{prl95p173901,optexpress14p6919}. Excellent agreement between
experiment and theory is apparent from Fig.~\ref{fig03} both in
the classical and quantum regime. It should be stressed that the
comparison to theory requires no adjustable parameters, and only
depends on the measured photon fluctuations of the incident light
source, the intensity transmission coefficient, and the total
collection efficiency of the setup.
\begin{figure}[ht]
  \centering
  \includegraphics[width=0.5\textwidth]{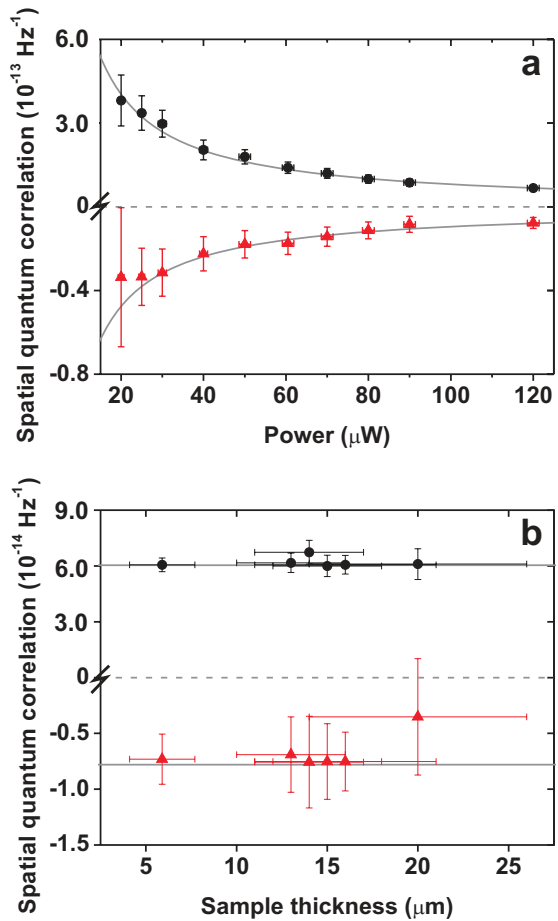}
\caption{(color online). (a) Measured spatial quantum correlation
function $\overline{C_{abab'}^{Q}}$ versus power of the incident
light beam. For classical (black points) and non-classical (red
triangles) photon fluctuations, positive and negative spatial
correlations are observed, respectively. Every data point
represents an average over three different positions on the sample
of thickness $L = 6\:\mu\text{m}$. The curves are the theoretical
predictions and the dashed line represents the uncorrelated case.
(b) Spatial quantum correlation function versus sample thickness
taken at an input power of $P = 120\:\mu\text{W}$. The spatial
quantum correlation function is found to be independent of the
sample thickness in agreement with theory (horizontal lines). }
  \label{fig04}
\end{figure}

The theory predicts that the multiple scattered photon
fluctuations depend on two terms varying as $L^{-1}$ and $L^{-2}$,
respectively, cf. Eq.~\eqref{eq03}. The latter term is only
observable for non vanishing spatial quantum correlations, i.e.
the comparison between experiment and theory allows extracting the
spatial quantum correlation function, cf. Fig.~\ref{fig04}. We
observe negative (positive) spatial correlations in the case where
the transmitted photon fluctuations are in the quantum (classical)
regime. According to Eq.~\eqref{eq02} the strength of the spatial
quantum correlation is expected to increase when reducing the
number of photons of the incident light. This pronounced behavior
is clearly demonstrated in Fig.~\ref{fig04}(a) and is in excellent
agreement with theory. Controlling the power of the non-classical
light source thus provides an efficient way of tuning the strength
of the spatial quantum correlations. Furthermore, the spatial
correlation function is predicted to be independent of the sample
thickness, which holds in the diffusive regime of multiple
scattering. This behavior is experimentally confirmed as well, see
Fig.~\ref{fig04}(b).

We presented the first experimental demonstration that photon
fluctuations below the classical limit can be observed after
multiple scattering using squeezed light as non-classical
resource. Multiple scattering was found to induce infinite range
spatial correlations that are of purely quantum origin. Our
experimental results were in excellent agreement with the full
quantum optics theory for multiple scattered light. We believe
that our results will inspire further explorations of the
multitude of new phenomena encoded in the quantum optical
properties of multiple scattered light.

We thank Jir\'i Janousek for help with the non-classical light
source and Elbert G. van Putten, Ivo M. Vellekoop, and Allard P.
Mosk for providing the samples. We gratefully acknowledge the
Danish Research Agency for financial support (project FNU
645-06-0503). A.H. and U.L.A. acknowledge support from the Danish Research Agency (project FTP 274-07-0509) and the EU project COMPAS for construction of the squeezed light source.

\clearpage


\begin{thebibliography}{}
\bibitem{note} Note that the spatial quantum correlation function
introduced here differs slightly from the one of Ref.
\cite{prl95p173901}. The two definitions contain the same physics
in the diffusive regime of consideration here, while they differ
in the mesoscopic regime.
\end{thebibliography}
\end{document}